\documentclass[preprint,3p]{elsarticle}

\usepackage{framed,multirow}

\usepackage{amssymb}
\usepackage{latexsym}

\usepackage{url}
\usepackage{xcolor}
\definecolor{newcolor}{rgb}{.8,.349,.1}
\usepackage{bm} % bold
\usepackage{physics}
\usepackage{amsmath}
\usepackage{amsthm}
\usepackage{mathtools}
\usepackage{booktabs}
\usepackage{yhmath}
\newcommand{\volume}{{\ooalign{\hfil$V$\hfil\cr\kern0.08em--\hfil\cr}}}
\usepackage{graphicx}
\usepackage[labelfont=it]{caption}
\usepackage{subcaption}
\usepackage{float}

\usepackage{cancel}
\usepackage[normalem]{ulem}

\usepackage{algorithm}
\usepackage{algpseudocode}
\usepackage{multicol}
\usepackage{tikz}

\usepackage{pifont}% http://ctan.org/pkg/pifont
\usepackage{hyperref}
\usepackage{cleveref}
\crefname{equation}{}{}
\usepackage{lineno}
\newcommand\patchAmsMathEnvironmentForLineno[1]{%
\expandafter\let\csname old#1\expandafter\endcsname\csname #1\endcsname
\expandafter\let\csname oldend#1\expandafter\endcsname\csname end#1\endcsname
\renewenvironment{#1}%
{\linenomath\csname old#1\endcsname}%
{\csname oldend#1\endcsname\endlinenomath}}%
\newcommand\patchBothAmsMathEnvironmentsForLineno[1]{%
\patchAmsMathEnvironmentForLineno{#1}%
\patchAmsMathEnvironmentForLineno{#1*}}%
\AtBeginDocument{%
\patchBothAmsMathEnvironmentsForLineno{equation}%
\patchBothAmsMathEnvironmentsForLineno{align}%
\patchBothAmsMathEnvironmentsForLineno{flalign}%
\patchBothAmsMathEnvironmentsForLineno{alignat}%
\patchBothAmsMathEnvironmentsForLineno{gather}%
\patchBothAmsMathEnvironmentsForLineno{multline}%
}
\theoremstyle{remark} 
\newtheorem{remark}{Remark}

\theoremstyle{plain}

\theoremstyle{plain}

\theoremstyle{definition}

\biboptions{sort&compress}

\journal{Computers \& Fluids}

\begin{document}

\begin{frontmatter}

\title{A Kinetic Energy Preserving and Entropy Conserving Two-Point Flux for Multi-species Compressible Flow}%
% \tnotetext[tnote1]{This is an example for title footnote coding.}
% Turbulent Flow Simulations using the Nonlinearly Stable Flux Reconstruction Method
\author[1]{Sai Shruthi Srinivasan\corref{cor1}}
\cortext[cor1]{Corresponding author.}

% \author[1]{Given-name2 \snm{Surname2}\fnref{fn1}}
% \fntext[fn1]{This is author footnote for second author.}  
% \author[2]{Given-name3 \snm{Surname3}}
%% Third author's email
% \ead{sai.srinivasan@mail.mcgill.ca}
\author[1]{Siva Nadarajah}
% \tnoteref{tnote1} % <-- This is an example for title footnote coding.
\address[1]{Department of Mechanical Engineering, McGill University, Montreal, Quebec H3A OC3, Canada}
% \address[2]{Department of Mechanical Engineering, McGill University, Montreal, Quebec H3A OC3, Canada}

% \received{1 May 2013}
% \finalform{10 May 2013}
% \accepted{13 May 2013}
% \availableonline{15 May 2013}
% \communicated{S. Sarkar}
%\begin{abstract}
%%%

%%%%
%\end{abstract}

%\begin{keyword}
%% Keywords
%\KWD Keyword1\sep Keyword2\sep Keyword3
%\end{keyword}

\end{frontmatter}
\vspace{-1.5cm}
%\linenumbers
\section{Introduction}\label{section: Introduction}
High-order entropy stable (ES) methods have attracted much interest in the computational fluid dynamics (CFD) research community over the past decade, as we seek to simulate practical engineering applications with high accuracy and robustness. As industry delves further into supersonic flight and space travel, we need schemes capable of modelling flows with multiple chemical species that interact and mix within the fluid. As such, there is a considerable amount of effort currently being focused on developing high-order ES schemes for multi-species (MS) flow \cite{ma2017entropy,gouasmi2020contributions,renac2021entropy,peyvan2023high,ching2024positivity,badrkhani2026entropy}. A key ingredient for these entropy stable schemes is the two-point flux, which determines the properties of the resulting high-order discretization. As such, the numerical scheme can possess important properties such as entropy conservation and kinetic energy preservation only if the two-point flux also mirrors these properties. {The importance of this is seen extensively in the single-species nonideal gas literature where the community is working towards fluxes that are not only entropy conserving and kinetic energy preserving, but also pressure equilibrium preserving as seen in the work of \citet{chan2026nodal} and \citet{coppola2026pressure}. In contrast, two-point fluxes used for high-order ES multi-species solvers in the literature are only entropy conserving and lack the kinetic energy preservation property.} The goal of this short communication is to develop a two-point flux that is both entropy conserving and kinetic energy preserving. To do so, we first start off by establishing the governing equations. For this work, we are considering the multi-species Euler equations \cite{giovangigli2012multicomponent}. These equations are related to the conservation of total mass, momentum, total energy and species mass.

\begin{equation}
    \bm{u} = \begin{pmatrix}\rho\\ \rho \mathbf{v}\\\rho e + \frac{1}{2}\rho \mathbf{v}^2\\ \rho_1\\\dots\\\rho_{s-1}\end{pmatrix},\quad \bm{f} = \begin{pmatrix}
        \rho \mathbf{v}\\ \rho \mathbf{v}^2 +P\\(\rho e + \frac{1}{2}\rho \mathbf{v}^2 + P)\mathbf{v}\\ \rho_1 \mathbf{v}\\ \dots \\ \rho_{N_s-1}\mathbf{v}
    \end{pmatrix},
\end{equation}
where $N_s$ denotes the number of species, $\rho_k$ is the species density, $\rho = \sum_{k=1}^{N_s} \rho_k$ is the mixture density, $\mathbf{v}$ is the fluid velocity. The pressure is determined using the ideal gas law:
\begin{equation}
    P = \sum_{k=1}^{N_s}\rho_kr_kT,\quad r_k = \frac{R}{m_k},
\end{equation}
where $m_k$ is the species molar mass and $R$ is the universal gas constant. To determine mixture internal energy, the polytropic ideal gas equations of state are used \cite{renac2021entropy}:
\begin{equation}\label{eq: internal-energy}
    P_k = \rho_ke_k(\gamma_k - 1), \quad e_k=c_{vk}T, \quad \rho e = \sum_{k=1}^{N_s}\rho_ke_k,
\end{equation}
where $e_k$ is the species specific internal energy and $c_{vk}$ is the constant volume specific heat. The internal energy is used to determine the temperature, $T$.

Other key relations used in this research include:
\begin{align}
    \text{Specific enthalpy of species: } &\quad h_k = e_k + r_kT\\ 
    \text{Species constant pressure specific heat: } &\quad c_{pk}=c_{vk}+r_k\\
    \text{Mass fraction of species: } &\quad Y_k=\frac{\rho_k}{\rho}\\
    \text{Specific heat ratio: } &\quad\gamma=\frac{c_p}{c_v} = \frac{\sum_k^{N_s}Y_kc_{pk}}{\sum_k^{N_s}Y_kc_{vk}}\\
    \text{Thermodynamic mixture entropy: } &\quad \rho s = \sum_{k=1}^s\rho_ks_k, \quad s_k = c_{vk}\ln{T} - r_k\ln(\rho_k)
\end{align}

In this work, we use these governing equations with the framework of generalized provably nonlinearly-stable flux reconstruction (NSFR) schemes introduced by \citet{cicchino2022nonlinearly}. The NSFR semidiscretization also requires the use of two-point fluxes. It was shown by \citet{tadmor1987numerical}, that the resulting discretization is entropy stable only if the two-point flux satisfies the entropy condition of \citet{harten1983symmetric}. This also applies to the NSFR scheme. Tadmor introduced a weak condition on the flux; referred to as the \textit{Tadmor shuffle condition}, it can be employed to derive an EC two-point flux \cite{tadmor1987numerical}. The shuffle condition is as follows :
\begin{align}
\label{eq:tadmorshuffle}
    [\bm{\eta}]^T \mathbf{f}^* = [\mathcal{F}],
\end{align}
where $\bm{\eta}$ denotes the vector of entropy variables, $\mathbf{f}^*$ denotes the EC flux and $\mathcal{F}$ denotes the entropy potential function. The entropy variables and the entropy potential function for multi-species compressible flow are \cite{gouasmi2020contributions,renac2021entropy}:
\begin{equation}
    \bm{\eta} = \frac{1}{T}\begin{bmatrix}
       g_{N_s} - \frac{\mathbf{v}^2}{2}\\
       \mathbf{v}\\
        -1\\
        g_1 - g_{N_s}\\
        \dots \\ 
        g_{N_s-1} - g_{N_s}\\
    \end{bmatrix}, \qquad \mathcal{F} = \sum_{k=1}^{N_s}\rho_kr_k\mathbf{v}
\end{equation}
\\
\begin{remark}
    Current entropy conserving two-point fluxes for multi-species \cite{gouasmi2020contributions,renac2021entropy,peyvan2023high,ching2024positivity, badrkhani2026entropy} are not kinetic energy preserving\\
\end{remark}

The standard two-point flux used for multi-species flow is the Chandrashekar ($CH$) flux \cite{gouasmi2020contributions} and it is derived using Eq.~\ref{eq:tadmorshuffle} and the following set of jump variables:
\begin{align}
    \bm{z} = \begin{bmatrix}
        \rho_1 & \dots & \rho_{N_s} & u & \frac{1}{T}
    \end{bmatrix}.
    \label{eq:ch_flux_z}
\end{align}

For a calorically perfect gas (CPG), the flux is given as:
\begin{align}
    \mathbf{f}^{\text{MS}_{\text{CH}}} = \begin{bmatrix}
         f_1\\ f_2\\ f_3\\ f_{4,1}\\ \dots \\ f_{4,s-1}
     \end{bmatrix} &= \begin{bmatrix}
         \sum_k^{N_s}\rho_{k}^{\ln}\bar{v}\\
         \frac{\sum_{k}^{N_s}r_k \bar{\rho}_{k}}{\overline{1/T}} + \bar{v}{f_{1}}\\
          \sum_k^{N_s}\left(e_{0,{k}} + \frac{c_{v,{k}}}{\left(\frac{1}{T}\right)^{\ln}} -\frac{1}{2}\overline{v^2}\right)\rho_{k}^{\ln}\bar{v} +\bar{v}{f_{2}}\\
          \rho_{1}^{\ln}\bar{v}\\
          \dots\\
          \rho_{N_s-1}^{\ln}\bar{v}          
     \end{bmatrix}
     \label{eq:ms-ch-flux}
\end{align}

Since it is derived using the Tadmor shuffle condition, it is an EC two-point flux. As per the literature, the $\text{MS}_{\text{CH}}$ flux is also considered to be KEP, but that is yet to be verified. Based on the \citet{jameson2008formulation} formulation, however, it would not be considered KEP, as the momentum flux component is not of the form:
\begin{align}
    f_{\rho v}^{num} = vf_{\rho}^{num} + \overline{P}.
    \label{eq:jameson-kep}
\end{align}

\begin{figure}[h!]
    \centering
    \includegraphics[width=0.49\linewidth]{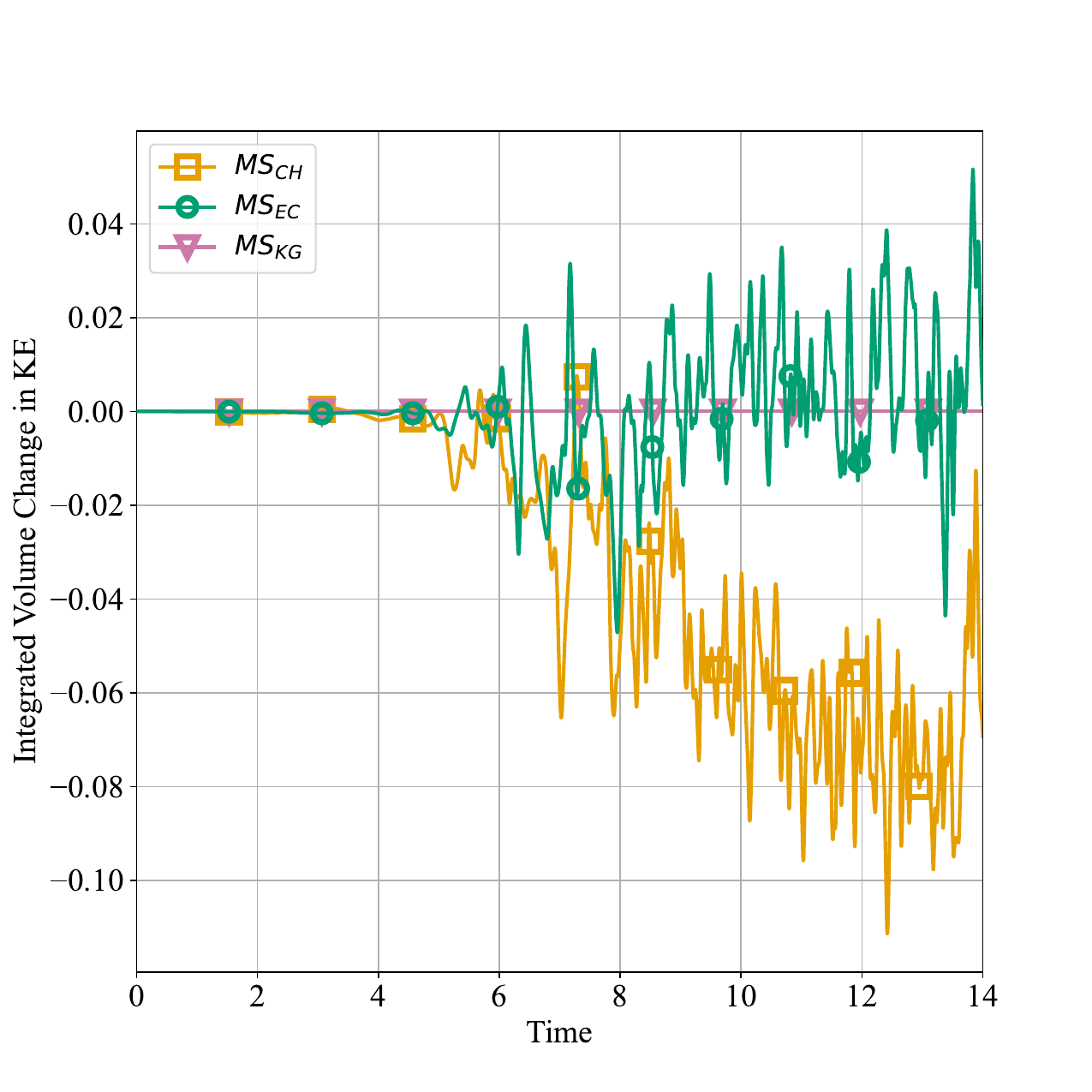}
    \includegraphics[width=0.49\linewidth]{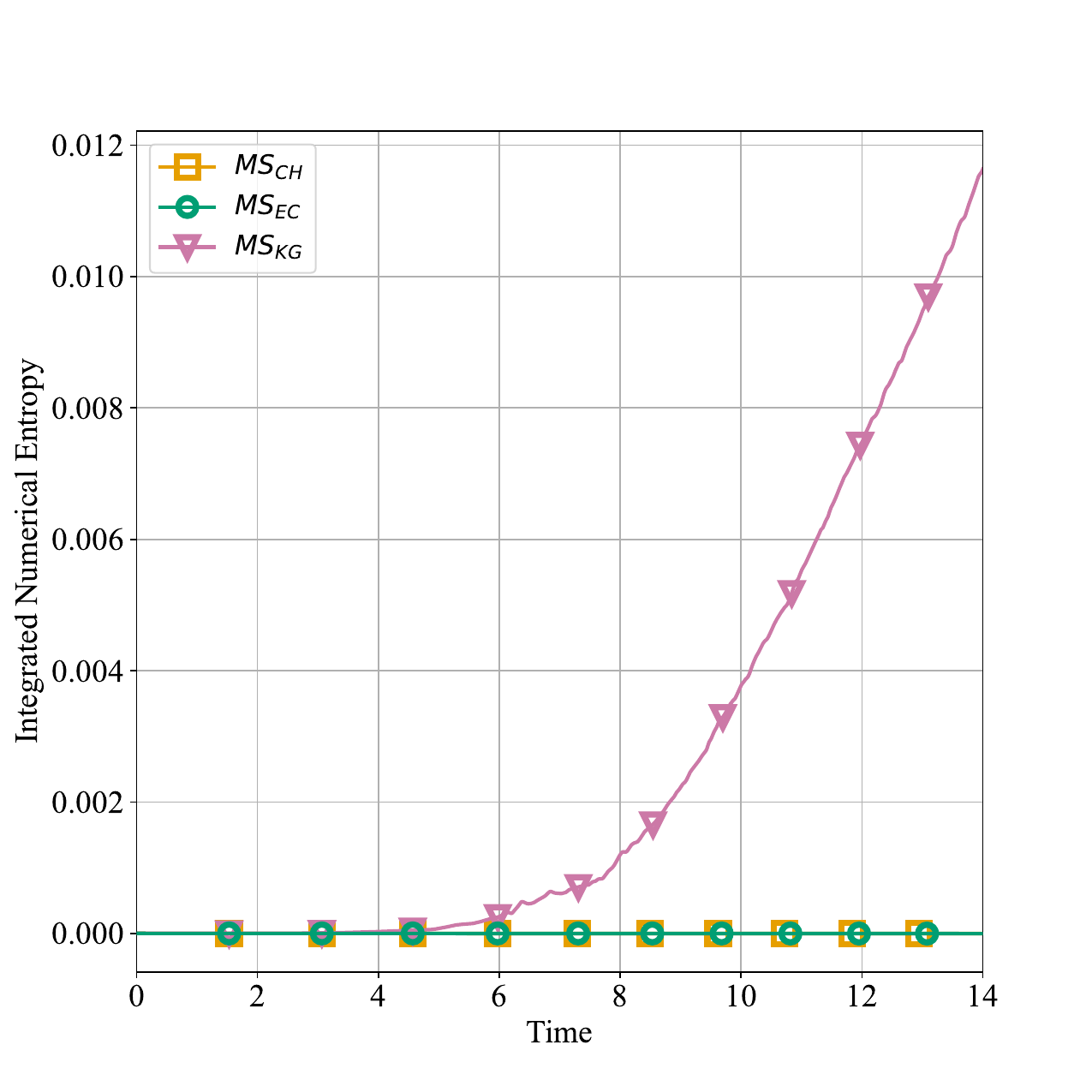}
    \caption{\textit{[Inviscid Taylor-Green Vortex]} This plot shows the results obtained for the inviscid TGV case using $p=3$, $8^3$ grid, GLL nodes and CFL$=0.1$. We compare the multi-species $CH$ flux outlined in Eq.~\ref{eq:ms-ch-flux} and the multi-species $KG$ flux outlined in Eq.~\ref{eq:ms-kg-flux}. The plot on the left shows the volume change in kinetic energy without pressure work, and the plot on the right shows the integrated entropy over time.}
    \label{fig:tgv-ch-ir-kg-only}
\end{figure}

Although the momentum component of the $CH$ flux includes a pressure term that resembles the arithmetic mean, it does not reduce to the arithmetic mean of the pressure:
\begin{align*}
    \frac{\sum_{k}^{N_s}r_k \bar{\rho}_{k}}{\overline{1/T}} &= \frac{\frac{1}{2}\sum_{k}^{N_s}r_k (\rho_{k,L}+\rho_{k,R})}{\frac{1}{2}\left(\frac{1}{T_L} + \frac{1}{T_R}\right)}\\
    &= \frac{\rho_Lr_L+\rho_Rr_R}{\frac{T_R + T_L}{T_LT_R}}\\
    &= \frac{(\rho_Lr_L+\rho_Rr_R)T_LT_R}{T_R + T_L} \neq \overline{P} = \frac{1}{2}(\rho_Lr_LT_L + \rho_Rr_RT_R)
\end{align*}
This is confirmed using the inviscid Taylor-Green vortex (TGV) test case similar to the numerical tests seen in \cite{gassner2016split,cicchino2025discretely}. The setup of the test case is provided in Sec.~\ref{sec: tgv} and the results for the $CH$ flux in comparison to other MS fluxes in current literature can be seen in Fig.~\ref{fig:tgv-ch-ir-kg-only}. For this case, we plot the volume change in kinetic energy without pressure work, as seen in \cite{cicchino2025discretely}, as well as the change in integrated numerical entropy over time. The $CH$ flux conserves entropy, but it does not preserve kinetic energy since it does not use the arithmetic mean of pressure in the momentum flux. As the $\text{MS}_{\text{CH}}$ flux does not possess the properties we desire, we must pursue other two-point fluxes.

Another popular option for the two-point flux in single-species simulations is the Ismail-Roe (IR) flux \cite{ismail2009affordable}. We can easily derive this EC flux for multi-species using similar jump variables as the single-species version:
\begin{align}
    z = \begin{bmatrix}
        \rho_1\sqrt{T} & \dots & \rho_s\sqrt{T} & \sqrt{\frac{1}{T}}v & \sqrt{\frac{1}{T}}
    \end{bmatrix}.
    \label{eq:ir_flux_z}
\end{align}
Using these variables and Eq.~\ref{eq:tadmorshuffle}, we derive an EC MS two-point flux that is similar to the single-species IR flux \cite{ismail2009affordable}:

\begin{align}
    \mathbf{f}^{\text{MS}_{\text{EC}}} = \begin{bmatrix}
         f_1\\ f_2\\ f_3\\ f_{4,1}\\ \dots \\ f_{4,s-1}
     \end{bmatrix} &= \begin{bmatrix}
         \sum_k^{s}\left(\rho_k\sqrt{T}\right)^{\ln}\overline{\left(\sqrt{\frac{1}{T}}v\right)}\\\\
         \frac{\sum_{k}^{s}r_k \overline{\rho_{k}\sqrt{T}}}{\overline{\sqrt{\frac{1}{T}}}} + \frac{\overline{\sqrt{\frac{1}{T}}v}}{\overline{\sqrt{\frac{1}{T}}}}{f_{1}}\\
          \sum_k^{s}\left(e_{0,{k}} + \frac{1}{\overline{\sqrt{\frac{1}{T}}}\left(\sqrt{\frac{1}{T}}\right)^{\ln}}(c_{v,{k}} + \frac{r_k}{2})\right)\left(\rho_k\sqrt{T}\right)^{\ln}\overline{\left(\sqrt{\frac{1}{T}}v\right)} +\frac{\overline{\sqrt{\frac{1}{T}}v}}{2\overline{\sqrt{\frac{1}{T}}}}{f_{2}}\\
          \left(\rho_1\sqrt{T}\right)^{\ln}\overline{\left(\sqrt{\frac{1}{T}}v\right)}\\
          \dots\\
          \left(\rho_{s-1}\sqrt{T}\right)^{\ln}\overline{\left(\sqrt{\frac{1}{T}}v\right)}        
     \end{bmatrix}
     \label{eq:ms-ir-flux}
\end{align}

%The $\text{MS}_{\text{EC}}$ flux is expected to be EC because of how it is derived (i.e., the Tadmor shuffle condition), but the momentum flux, again, does not include the arithmetic mean of pressure, so it is not expected to be KEP. 
The $\text{MS}_{\text{EC}}$ flux is expected to satisfy the entropy conservation (EC) property by construction, as it is derived from the Tadmor shuffle condition. In contrast, its momentum flux does not include the arithmetic mean of the pressure, and therefore the flux is not expected to be kinetic energy preserving (KEP).
This is proven using the inviscid TGV test, where the plot of kinetic energy without pressure work in Fig.~\ref{fig:tgv-ch-ir-kg-only} shows that the $\text{MS}_{\text{EC}}$ flux does not preserve kinetic energy, and it also has a similar magnitude of variation in KE as the $\text{MS}_{\text{CH}}$ flux. The $\text{MS}_{\text{EC}}$ flux could also potentially have robustness issues due to the influence of temperature on the density flux, similar to the single-species version, where discontinuities in pressure lead to the mass flux being overestimated and nonphysical, which leads to the scheme breaking down as explained by \citet{derigs2017novel}.

Another option for the two-point flux that uses Eq.~\ref{eq:tadmorshuffle} is that of \citet{renac2021entropy}. However, we have chosen to limit the study of the EC fluxes to the $\text{MS}_{\text{CH}}$ and $\text{MS}_{\text{EC}}$ fluxes, as our implementation does not use the assumption of mechanical equilibrium.
\begin{remark}
    The current option for a kinetic energy preserving two-point flux for multi-species, ie. the Pirozzoli flux \cite{keeton2025discontinuous}, is not entropy conserving.\\
\end{remark}
If we look to the literature for KEP fluxes, the Pirozzoli flux employed by \citet{keeton2025discontinuous} is an apparent choice. The Pirozzoli flux has the crucial KEP property since the momentum flux has the form given in Eq.~\ref{eq:jameson-kep}, but it lacks the EC property that is also of interest. In this work, we will use the equivalent Kennedy-Gruber (KG) flux \cite{kennedy2008reduced} to compare the properties of the $\text{MS}_{\text{CH}}$ flux and the $\text{MS}_{\text{EC}}$ two-point flux. For calorically perfect gas, the $\text{MS}_{\text{KG}}$ flux is given as:

\begin{align}
    \mathbf{f}^{\text{MS}_{\text{KG}}} = \begin{bmatrix}
         f_1\\ f_2\\ f_3\\ f_{4,1}\\ \dots \\ f_{4,s-1}
     \end{bmatrix} &= \begin{bmatrix}
         \sum_k^{N_s}\overline{\rho_{k}}\bar{v}\\
         \overline{P} + \bar{v}{f_{1}}\\
          \overline{\rho}(\overline{e} + \overline{P})\overline{v}\\
          \overline{\rho_{1}}\bar{v}\\
          \dots\\
          \overline{\rho}_{N_s-1}\bar{v}          
     \end{bmatrix}
     \label{eq:ms-kg-flux}
\end{align}

We can see that the $\text{MS}_{\text{KG}}$ flux has the arithmetic mean of pressure in the momentum flux and thus is expected to be KEP. Since it is not derived with the Tadmor condition, the two-point flux is not EC. Referring once more to the inviscid TGV test case, we see in Fig.~\ref{fig:tgv-ch-ir-kg-only} that the $\text{MS}_{\text{KG}}$ flux is indeed KEP, but it is not EC.

The results presented in Fig.~\ref{fig:tgv-ch-ir-kg-only} show that the change in kinetic energy for the EC fluxes ($\text{MS}_{\text{CH}}$ and $\text{MS}_{\text{EC}}$) is not at machine precision, and it is several orders of magnitude larger than the change seen in the $\text{MS}_{\text{KG}}$ flux. The lack of KEP is also observed for the single-species $CH$ and $IR$ fluxes \cite{gassner2016split,ranocha2018generalised}. This KEP property is critical for accurate simulations of turbulence, as there is an energy cascade between the different scales \cite{jameson2008formulation}. As for entropy conservation, we observe that the $\text{MS}_{\text{KG}}$ flux is not EC and has a change in entropy that is several orders of magnitude greater than the $\text{MS}_{\text{CH}}$ and $\text{MS}_{\text{EC}}$ fluxes. 

%This leads us to the objective of this short communication. To be able to accurately simulate multi-species compressible flows, we require a two-point flux that preserves kinetic energy and conserves entropy, so in this short communication we will use a process similar to that of \citet{ranocha2018generalised} to derive an EC, KEP two-point flux for calorically and thermally perfect multi-species flow.
This motivates the objective of the present short communication. Accurate simulation of multi-species compressible flows requires a two-point flux that is both entropy conservative and kinetic energy preserving. {Accordingly, following an approach similar to the single-species strategy used by \citet{ranocha2018generalised} for calorically perfect gas and the one used by \citet{aiello2026formulation} for thermally perfect gas, we derive an EC-KEP two-point flux for calorically and thermally perfect multi-species flows. Additionally, we demonstrate the performance of the EC-KEP flux numerically for multi-species flow compared to the MS fluxes introduced in Eq.\Cref{eq:ms-ch-flux,eq:ms-ir-flux,eq:ms-kg-flux}.}
\section{An Entropy Conserving and Kinetic Energy Preserving Two-Point Flux for Multi-Species Flows}\label{sec: flux-derivation}
As the $\text{MS}_{\text{CH}}$ flux does not adhere to the formulation required for KEP, there is a need for a flux that is both EC and KEP. For the single-species formulation, \citet{ranocha2018generalised} derives a flux with the same jump variables as the CH flux, but enforces KEP by selecting a momentum flux that satisfies the Jameson formulation \cite{jameson2008formulation} and derives the energy component of the flux using the fixed density and momentum flux. We follow a similar process to obtain a EC and KEP two-point flux for multi-species flow. 

As per the Jameson formulation, the fixed components of the flux are:
\begin{align*}
    f_{\rho,k}^{num} &= \rho_k^{\ln}\bar{v}\\
    f_{\rho}^{num} &= \sum_{k}^{N_s}  \rho_k^{\ln}\bar{v}\\
    f_{\rho v}^{num} &= \bar{v}f_{\rho}^{num} + \overline{P}
\end{align*}

The only remaining component is the energy flux. To obtain the energy flux, we use the same jump variables as the $\text{MS}_{\text{CH}}$ flux, given in Eq.~\ref{eq:ch_flux_z} and the Tadmor shuffle condition given in Eq.~\ref{eq:tadmorshuffle}:
\begin{align*}
    [\bm{\eta}]^T \mathbf{f}^* &= [\mathcal{F}]\\
    0 &= [\bm{\eta}]^T \mathbf{f}^* - [\mathcal{F}]\\
    0 &= \sum_k^{N_s-1}[z_{1,k}]\frac{r_k}{z_{1,k}^{\ln}}{f_{4,k}} + [z_{1,N_s}]\left(\frac{r_{N_s}}{z_{1,N_s}^{\ln}}{f_{1}} - \sum_k^{N_s-1}\frac{r_{N_s}}{z_{1,N_s}^{\ln}}{f_{4,k}}\right) + [z_2]\left(\bar{z}_3{f_{2}}-\bar{z}_2\bar{z}_3{f_{1}}\right)\\
     &\quad+[z_3] \left(\left(e_{0,N_s} + \frac{c_{v,N_s}}{z_3^{\ln}} - \frac{1}{2}\bar{z_2^2}\right){f_{1}} + \bar{z}_2{f_{2}} - {f_{3}} + \sum_k^{N_s-1}\left(e_{0,{k}} - e_{0,N_s} + \frac{c_{v,{k}}}{z_3^{\ln}} - \frac{c_{v,N_s}}{z_3^{\ln}}\right){f_{4,k}}\right)\\
     &\quad- \sum_{k=1}^{N_s} r_k \bar{z}_{1,k} [z_2] - \sum_{k=1}^{N_s-1} r_k [z_{1,k}] \bar{z}_2 - r_{N_s}[z_{1,N_s}]\bar{z}_2\\
\end{align*}

Inserting our fixed flux terms into the equation, we can derive our new energy flux:
\begin{align*}
     0 &= \sum_k^{N_s-1}[\rho_{k}]\frac{r_k}{\rho_{k}^{\ln}}{\left(\rho_k^{\ln}\bar{v}\right)} + [\rho_{N_s}]\left(\frac{r_{N_s}}{\rho_{N_s}^{\ln}}{\left(\sum_{k}^{N_s}  \rho_k^{\ln}\bar{v}\right)} - \sum_k^{N_s-1}\frac{r_{N_s}}{\rho_{N_s}^{\ln}}{\left(\rho_k^{\ln}\bar{v}\right)}\right)\\
     &\quad+ [v]\left(\overline{\frac{1}{T}}{\left(\bar{v}\left(\sum_{k}^{N_s}  \rho_k^{\ln}\bar{v}\right) + \overline{P}\right)}-\bar{v}\overline{\frac{1}{T}}{\left(\sum_{k}^{N_s}  \rho_k^{\ln}\bar{v}\right)}\right)\\
     &\quad+\left[\frac{1}{T}\right] \left(\left(e_{0,N_s} + \frac{c_{v,N_s}}{\left(\frac{1}{T}\right)^{\ln}} - \frac{1}{2}\overline{v^2}\right){\left(\sum_{k}^{N_s}  \rho_k^{\ln}\bar{v}\right)} + \bar{v}{\left(\bar{v}\left(\sum_{k}^{N_s}  \rho_k^{\ln}\bar{v}\right) + \overline{P}\right)} - {f_{3}}\right)\\ &\quad+ \left[\frac{1}{T}\right]\left(\sum_k^{N_s-1}\left(e_{0,{k}} - e_{0,N_s} + \frac{c_{v,{k}}}{\left(\frac{1}{T}\right)^{\ln}} - \frac{c_{v,N_s}}{\left(\frac{1}{T}\right)^{\ln}}\right){\left(\rho_k^{\ln}\bar{v}\right)}\right)- \sum_{k=1}^{N_s} r_k \bar{\rho}_{k} [v] - \sum_{k=1}^{N_s-1} r_k [\rho_{k}] \bar{v} - r_{N_s}[\rho_{N_s}]\bar{v}\\
     0 &= \sum_k^{N_s-1}[\rho_{k}]r_k\bar{v} + [\rho_{N_s}]r_{N_s}\overline{v} + \overline{P}\overline{\frac{1}{T}}[v]+\left[\frac{1}{T}\right]\overline{v}\left(\sum_k^{N_s}\left(e_{0,k}+\frac{c_{v,k}}{\left(\frac{1}{T}\right)^{\ln}} - \frac{1}{2}\overline{v^2} + (\overline{v})^2\right)\rho_k^{\ln} + \overline{P}\right)\\
     &\quad- \sum_{k=1}^{N_s} r_k \bar{\rho}_{k} [v] - \sum_{k=1}^{N_s-1} r_k [\rho_{k}] \bar{v} - r_{N_s}[\rho_{N_s}]\bar{v} - \left[\frac{1}{T}\right]{f_{3}}\\
\end{align*}

Isolating the energy term provides us with the following flux for total energy:
\begin{align}
     {f_{3}} &= {f_{\rho E}^{num}} = \overline{v}\left(\sum_k^{N_s}\left(e_{0,k}+\frac{c_{v,k}}{\left(\frac{1}{T}\right)^{\ln}} - \frac{1}{2}\overline{v^2} + (\overline{v})^2\right)\rho_k^{\ln} + \overline{P}\right) + {\left(\overline{P}\overline{\frac{1}{T}} - \sum_k^{N_s}r_k\overline{\rho_k}\right)\frac{[v]}{\left[\frac{1}{T}\right]}}
\end{align}

We now have our energy component for the KEP and EC flux. The last term can be further simplified:
\begin{align*}
    {\left(\overline{P}\overline{\frac{1}{T}} - \sum_k^{N_s}r_k\overline{\rho_k}\right)\frac{[v]}{\left[\frac{1}{T}\right]}} &= \left(\overline{P}\overline{\frac{1}{T}} - \sum_k^{N_s}r_k\overline{\rho_k}\right)\frac{[v]}{\left[\frac{1}{T}\right]} \\
    &= \left(\frac{1}{2}\left(\rho_Lr_LT_L + \rho_Rr_RT_R\right)\frac{1}{2}\left(\frac{1}{T_L}+\frac{1}{T_R}\right) - \sum_k^{N_s}\frac{1}{2}(\rho_{k,L}+\rho_{k,R})r_k\right)\frac{[v]}{\left[\frac{1}{T_L} - \frac{1}{T_R}\right]}\\
    &= \left(\frac{1}{4}\left(\rho_Lr_LT_L + \rho_Rr_RT_R\right)\left(\frac{T_R+T_L}{T_LT_R}\right) - \frac{1}{2}\rho_Lr_L-\frac{1}{2}\rho_Rr_R\right)T_LT_R\frac{[v]}{\left[T_R-T_L\right]}\\
    &= \frac{1}{4}\left(\frac{\rho_Lr_LT_L}{T_R} + \frac{\rho_Rr_RT_R}{T_L} - \rho_Lr_L - \rho_Rr_R\right)T_LT_R\frac{[v]}{T_R-T_L}\\
    &= \frac{1}{4}\left(\rho_Lr_LT_L^2 -\rho_Lr_LT_LT_R + \rho_Rr_RT_R^2 - \rho_Rr_RT_LT_R\right)\frac{[v]}{T_R-T_L}\\
    &= \frac{1}{4}\left(-\rho_Lr_LT_L(T_R-T_L) + \rho_Rr_RT_R(T_R-T_L)\right)\frac{[v]}{T_R-T_L}\\
    &= -\frac{1}{4} (\rho_Lr_LT_L - \rho_Rr_RT_R)[v] = -\frac{1}{4}(P_L-P_R)[v] = {-\frac{1}{4}[P][v]}
\end{align*}

This yields the final form of our flux:
\begin{align}
    \mathbf{f}^{\text{MS}_{\text{EC/KEP}}} = \begin{bmatrix}
         f_1\\ f_2\\ f_3\\ f_{4,1}\\ \dots \\ f_{4,s-1}
     \end{bmatrix} &= \begin{bmatrix}
         \sum_k^{N_s}\rho_{k}^{\ln}\bar{v}\\
         \bar{v}{f_{1}} + \overline{P}\\
          \sum_k^{N_s}\left(e_{0,{k}} + \frac{c_{v,{k}}}{\left(\frac{1}{T}\right)^{\ln}} -\frac{1}{2}\overline{v^2} + (\bar{v})^2\right)\rho_{k}^{\ln}\bar{v} +\overline{P}\bar{v} { - \frac{1}{4}[P][v]}\\
          \rho_{1}^{\ln}\bar{v}\\
          \dots\\
          \rho_{N_s-1}^{\ln}\bar{v}          
     \end{bmatrix}
\end{align}
This fix is similar to the single-species CPG Ranocha pressure fix for the Chandrashekar flux \cite{ranocha2018generalised}  { as well as the MS-CPG generalization derived in \cite{aiello2026formulation}.}

\subsection{Thermally Perfect Gas Flux}
The process of deriving a KEP flux only majorly affects the momentum components of the $\text{MS}_{\text{CH}}$ flux, as we see in the calorically perfect version. This change can be applied directly to the TPG version without any extra derivations:
\begin{align}
    \mathbf{f}^{\text{MS}_{\text{EC/KEP}}, \text{TPG}} = \begin{bmatrix}
         f_1\\ f_2\\ f_3\\ f_{4,1}\\ \dots \\ f_{4,s-1}
     \end{bmatrix} &= \begin{bmatrix}
         \sum_k^{s}\rho_{k}^{\ln}\bar{v}\\
         \bar{v}{f_{1}} + \overline{P}\\
          \sum_k^{s}\left(h_{k,ref} + \frac{a_{0,k}-r_k}{(1/T)^{ln}} + \sum_{p=1}^5\frac{a_{p,k}}{p(p+1)}(f_p(T)T^\times) - \frac{1}{2}\overline{v^2}\right)\rho_{k}^{\ln}\bar{v} +\overline{P}\bar{v} {- \frac{1}{4}[P][v]}\\
          \rho_{1}^{\ln}\bar{v}\\
          \dots\\
          \rho_{s-1}^{\ln}\bar{v}          
     \end{bmatrix}
\end{align}
where $h_{k,ref} = h_{k,0}-h_k(T_{ref})$ and $f_p(T)$ is an averaging operator that is given as
\begin{align*}
    f_1(T) &= 1\\
    f_2(T) &= 2\overline{T}\\
    f_3(T) &= \overline{T^2} + 2\overline{T}\overline{T}\\ 
    f_4(T) &= 4\overline{T^2}\overline{T}\\ 
    f_5(T) &= 2\overline{T^3}\overline{T} + \overline{T^2}\overline{T^2} + 2\overline{T^2}\overline{T}\overline{T}. 
\end{align*}
For further details on the averaging operator, the reader is referred to \cite{gouasmi2020contributions} and \cite{ching2024positivity}. {The extension to TPG presented here is also similar to the MS-TPG generalization in \cite{aiello2026formulation}.}
\section{Numerical Results}\label{sec: results}
The performance of the newly proposed two-point flux is demonstrated using two different test cases in this section. All the tests use the nonlinearly-stable flux reconstruction semi-discretization of Cicchino et al.~\cite{cicchino2022nonlinearly,cicchino2025discretely}, and Gauss-Lobatto-Legendre (GLL) nodes are used for the solution and flux bases. The time advancement method of choice is the strong stability preserving third-order accurate explicit Runge-Kutta (SSPRK3) method. The convective-based CFL condition used in this work is determined as follows:
\begin{align}
    \Delta t = CFL \: \frac{\Tilde{\Delta x}}{\lambda_{\max}}, \quad \Tilde{\Delta x} = \frac{x_{\max}-x_{\min}}{(DOF)^{1/dim}}, \quad \lambda_{\max} = \max(|v|+c),
\end{align}
where $\Tilde{\Delta x}$ is the approximate grid spacing and $\lambda_{max}$ is the maximum wavespeed of the initial condition with speed of sound $c = \sqrt{\gamma R T}$.

\subsection{Density Pulse Advection}\label{sec: density-pulse}
The 1D density pulse test case introduced by \citet{wang2019partial} involves the advection of a gas mixture consisting of multiple species and has a constant initial velocity and pressure. 
The initialization for this test case is:
\begin{align*}
    T &= T_0 - \frac{(\gamma_0-1)\Gamma^2}{8\gamma_0\pi}\exp{\left(\frac{1-r^2}{2}\right)}\\
    Y_{H_2} &= Y_{H_2,0} - \frac{2\pi a_1}{\gamma_0\Gamma}\exp{\left(\frac{1-r^2}{2}\right)}\\
    Y_{O_2} &= 1.0- Y_{H_2}\\
    P&= P_0 = 101325 \text{ Pa}\\
    v_x &= v_{0} = 100 \text{ m/s},
\end{align*}
where $T_0 = 300$K, $\Gamma = 50$, $\gamma_0 = 1.4$, $Y_{H_2,0} = 0.01277$, $a_1=0.005$ and $r=\sqrt{(x-x_0)^2}$ (with $x_0=5.0$).\\

\begin{figure}[h!]
    \centering
    \includegraphics[width=0.41\linewidth]{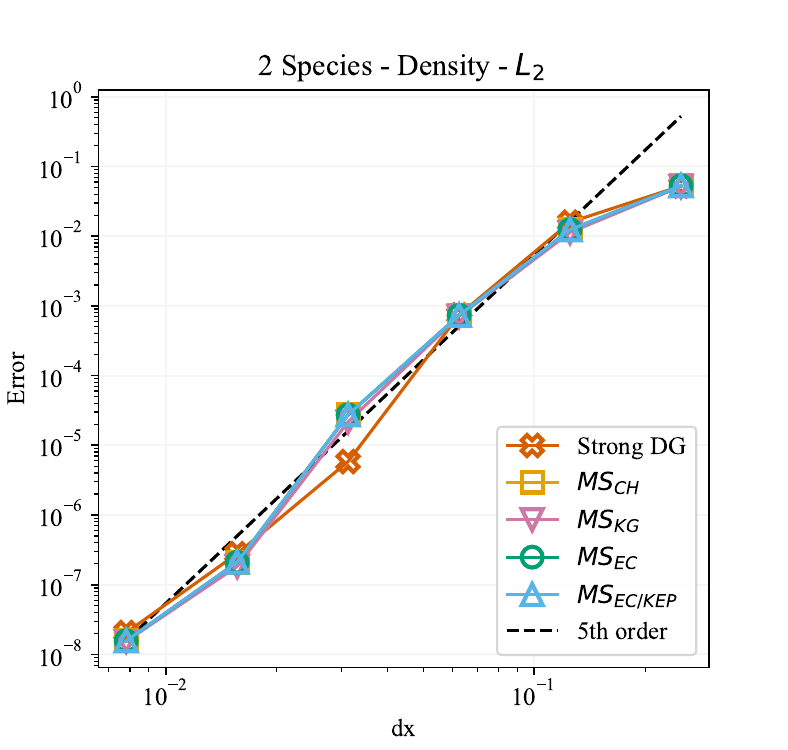}
    \includegraphics[width=0.41\linewidth]{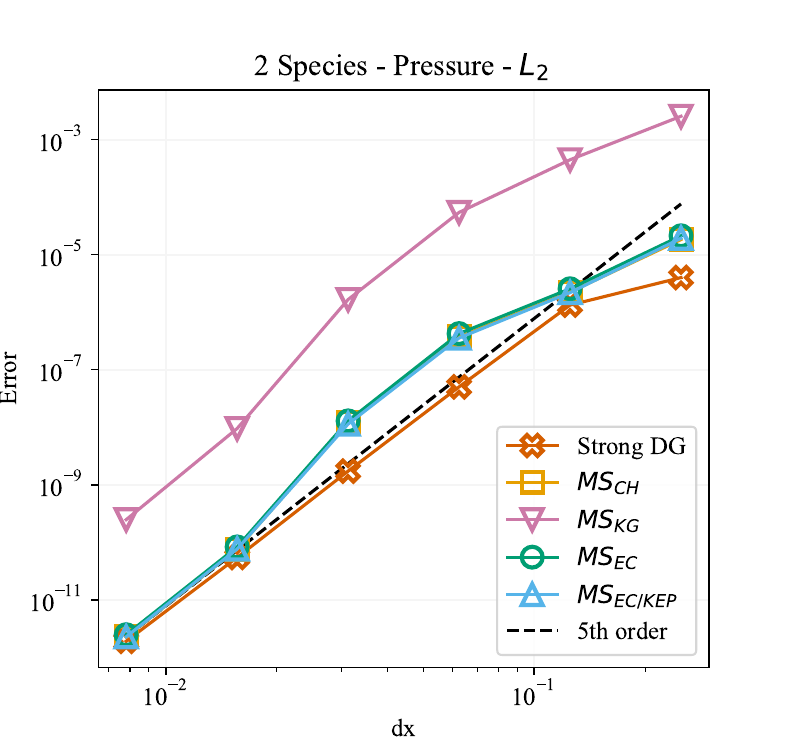}
    \includegraphics[width=0.41\linewidth]{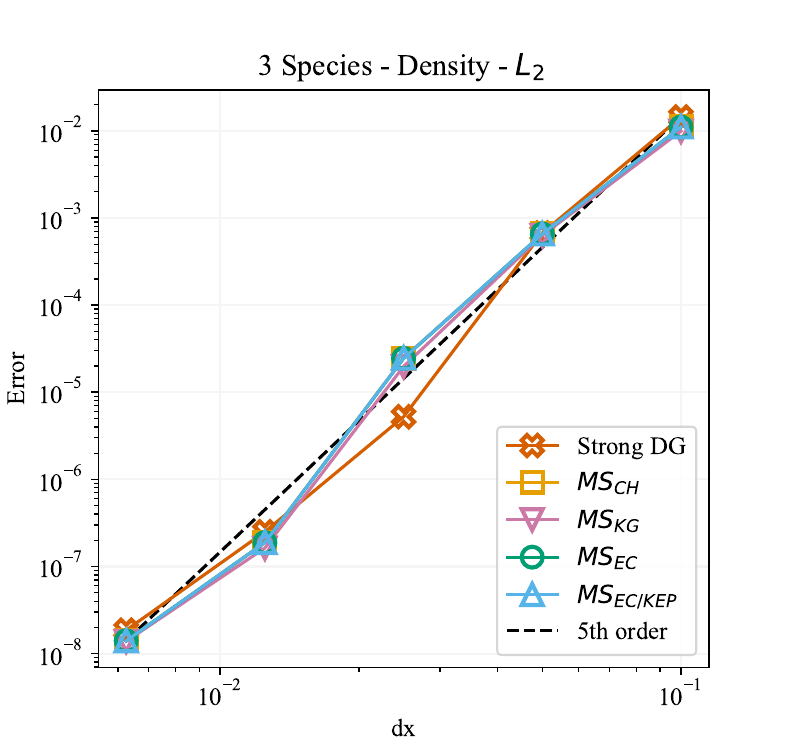}
    \includegraphics[width=0.41\linewidth]{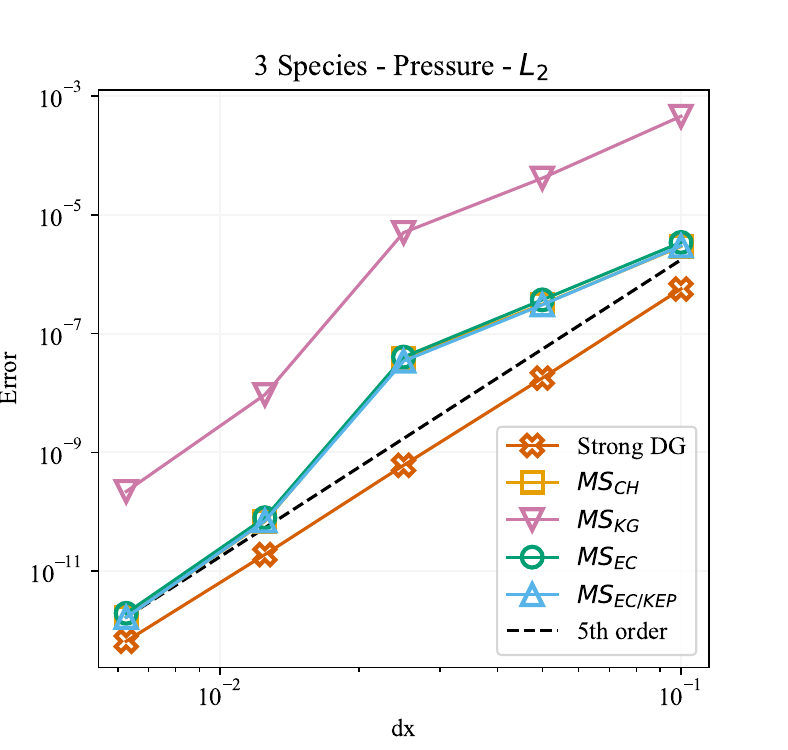}
    \includegraphics[width=0.41\linewidth]{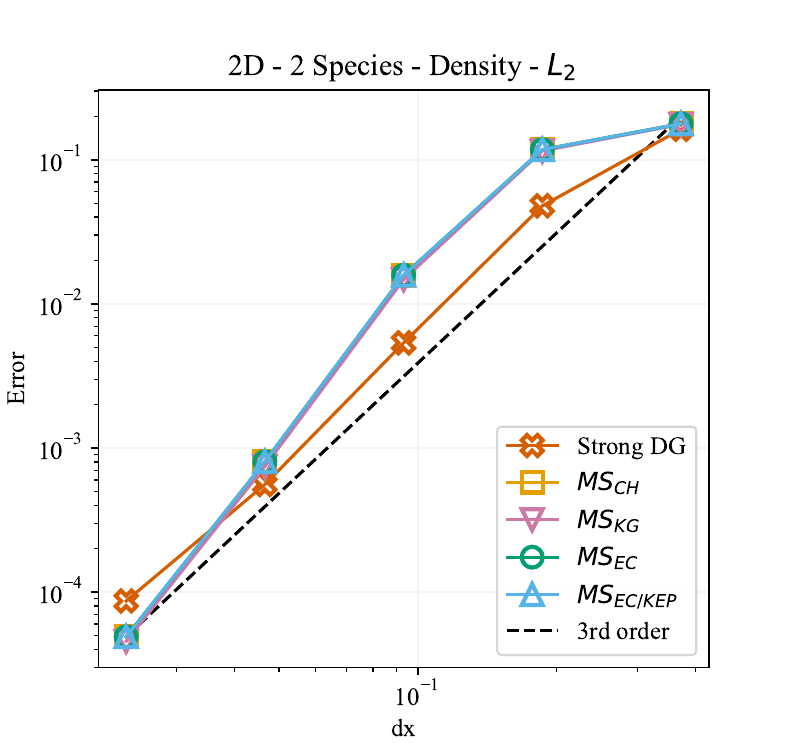}
    \includegraphics[width=0.41\linewidth]{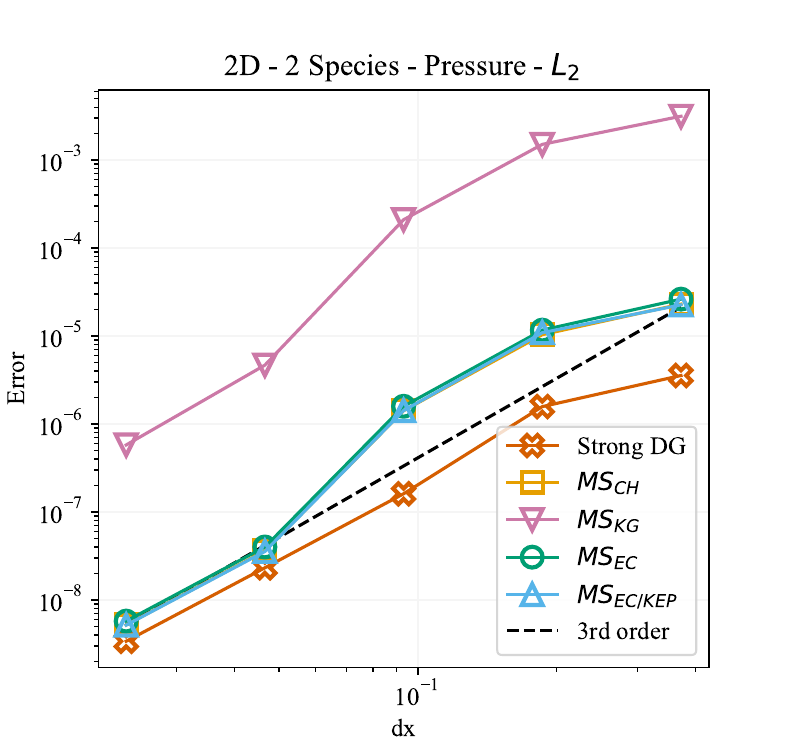}
    \caption{\textit{[Density Pulse Test]} Grid convergence orders with GLL flux nodes are plotted for density and pressure. The top row is the 2-species configuration, and the middle row is the 3-species configuration, both of which use $p4$ and CFL$=0.25$. The bottom row is the 2 species, 2D configuration which uses $p2$ and CFL$=0.1$.}
    \label{fig:density-pulse-grid-ooa}
\end{figure}

For the 3-species configuration, the mass fractions are initialized as:
\begin{align*}
    Y_{H_2} &= Y_{H_2,0} - \frac{2\pi a_1}{\gamma_0\Gamma}\exp{\left(\frac{1-r^2}{2}\right)}\\
    Y_{O_2} &= Y_{O_2,0} - \frac{2\pi a_2}{\gamma_0\Gamma}\exp{\left(\frac{1-r^2}{2}\right)}\\
    Y_{N_2} &= 1.0- (Y_{H_2}+Y_{O_2}),
\end{align*}
with $Y_{O_2,0} = 0.101$. For the 2D 2-species configuration, we use $r=\sqrt{(x-x_0)^2+(y-y_0)^2}$ (with $x_0=y_0=5.0$) and the velocities are initialized as:
\begin{align*}
    v_x &= v_{0} = 100 \text{ m/s}\\
    v_y &= v_{0} = 100 \text{ m/s}.
\end{align*}
This test case uses a domain of $[0,10]$ for 1D and $[0,10]^2$ for 2D. The final time is set such that the pulse advects for one full cycle and returns to the initial condition. The density and pressure orders of convergence for Strong-DG, $\text{MS}_{\text{KG}}$, $\text{MS}_{\text{CH}}$, $\text{MS}_{\text{EC}}$, and the proposed $\text{MS}_{\text{EC/KEP}}$ are shown in Fig.~\ref{fig:density-pulse-grid-ooa} for all three configurations. All schemes achieve the expected convergence orders. In all configurations, the $\text{MS}_{\text{KG}}$ flux has a noticeably greater magnitude in error for pressure compared to the other four schemes. The performance of the $\text{MS}_{\text{EC/KEP}}$ flux is equivalent to the $\text{MS}_{\text{CH}}$ flux as expected. The $\text{MS}_{\text{EC}}$ flux also achieves similar orders as the other EC fluxes. The two-point fluxes are essentially dissipation-free since we do not use an interface stabilisation term. On the other hand, the Strong DG scheme has dissipation from the Lax-Friedrichs flux and is expected to achieve better orders on coarser grids, especially for pressure, as it suppresses the spurious pressure oscillation phenomena that is common in multi-species simulation.\\

In addition to the OOA results, entropy change at the final time is compared for Strong-DG, $\text{MS}_{\text{CH}}$, $\text{MS}_{\text{EC}}$ and $\text{MS}_{\text{EC/KEP}}$ using time step refinement similar to the analysis shown in \cite{renac2021entropy}. The results in Table~\ref{tab:density-pulse-time-ooa} demonstrate that the change in entropy converges to machine precision for $\text{MS}_{\text{CH}}$, $\text{MS}_{\text{EC}}$ and $\text{MS}_{\text{EC/KEP}}$ with design order 3 as the time step is refined, which corresponds with the SSPRK3 method. This confirms that the two-point fluxes are indeed entropy conservative.

\begin{table}[h!]
\centering
\small
\caption{\textit{[Density Pulse Test]} Timestep Convergence Study of Entropy with Strong DG - Lax Friedrichs, $\text{MS}_{\text{CH}}$, $\text{MS}_{\text{EC}}$ and $\text{MS}_{\text{EC/KEP}}$ flux, p4, GLL flux nodes and N=4 elements}
\label{tab:density-pulse-time-ooa}
\begin{tabular}{lcccccccc}
\hline
\multicolumn{1}{c}{}           & \multicolumn{2}{c}{Strong DG} & \multicolumn{2}{c}{$\text{MS}_{\text{CH}}$}                  & \multicolumn{2}{c}{$\text{MS}_{\text{EC}}$}             & \multicolumn{2}{c}{$\text{MS}_{\text{EC/KEP}}$}        \\ \hline
\multicolumn{1}{c}{$\Delta t$} & $|(\rho s) - (\rho s)_0|$    & OOA             & $|(\rho s) - (\rho s)_0|$ & OOA           & $|(\rho s) - (\rho s)_0|$ & OOA           & $|(\rho s) - (\rho s)_0$ & OOA           \\ \hline
0.1                            & 1.0804e-3  & -        & 2.8178e-6  & -        & 2.8166e-06  & -       & 2.8179e-06  & -             \\
0.05                           & 1.0784e-3  & 0.0026   & 3.5303e-7  & 2.9967   & 3.5302e-07  & 2.9961  & 3.5303e-07  & 2.9967 \\
0.025                          & 1.0781e-3  & 0.0004   & 4.4153e-8  & 2.9992   & 4.4133e-08  & 2.9998  & 4.4153e-08  & 2.9992  \\
0.0125                         & 1.0781e-3  & 0.00     & 5.5372e-9  & 2.9953   & 5.5167e-09  & 2.9999  & 5.5378e-09  & 2.9951 \\
0.00625                        & 1.0781e-3  & 0.00     & 7.0923e-10 & 2.9648   & 6.8977e-10  & 2.9996  & 7.0983e-10  & 2.9637 \\
0.003125                       & 1.0781e-3  & 0.00     & 1.0561e-10 & 2.7475   & 8.6231e-11  & 2.9998  & 1.0650e-10  & 2.7366 \\ \hline
% 0.0015625                      & 1.0781e-3  & 0.00     & 3.0240e-11 & 1.8042   & 1.0885e-11  & 2.9858  & 3.0610e-11  & 1.7987 \\
% 0.00078125                     & 1.0781e-3  & 0.00     & 2.0378e-11 & 0.5695   & 1.1084e-12  & 3.2958  & 2.1600e-11  & 0.5029 \\ \hline
\end{tabular}
\end{table}
\subsection{Inviscid Taylor-Green Vortex}\label{sec: tgv}
\begin{figure}[h!]
    \centering
    \includegraphics[width=0.49\linewidth]{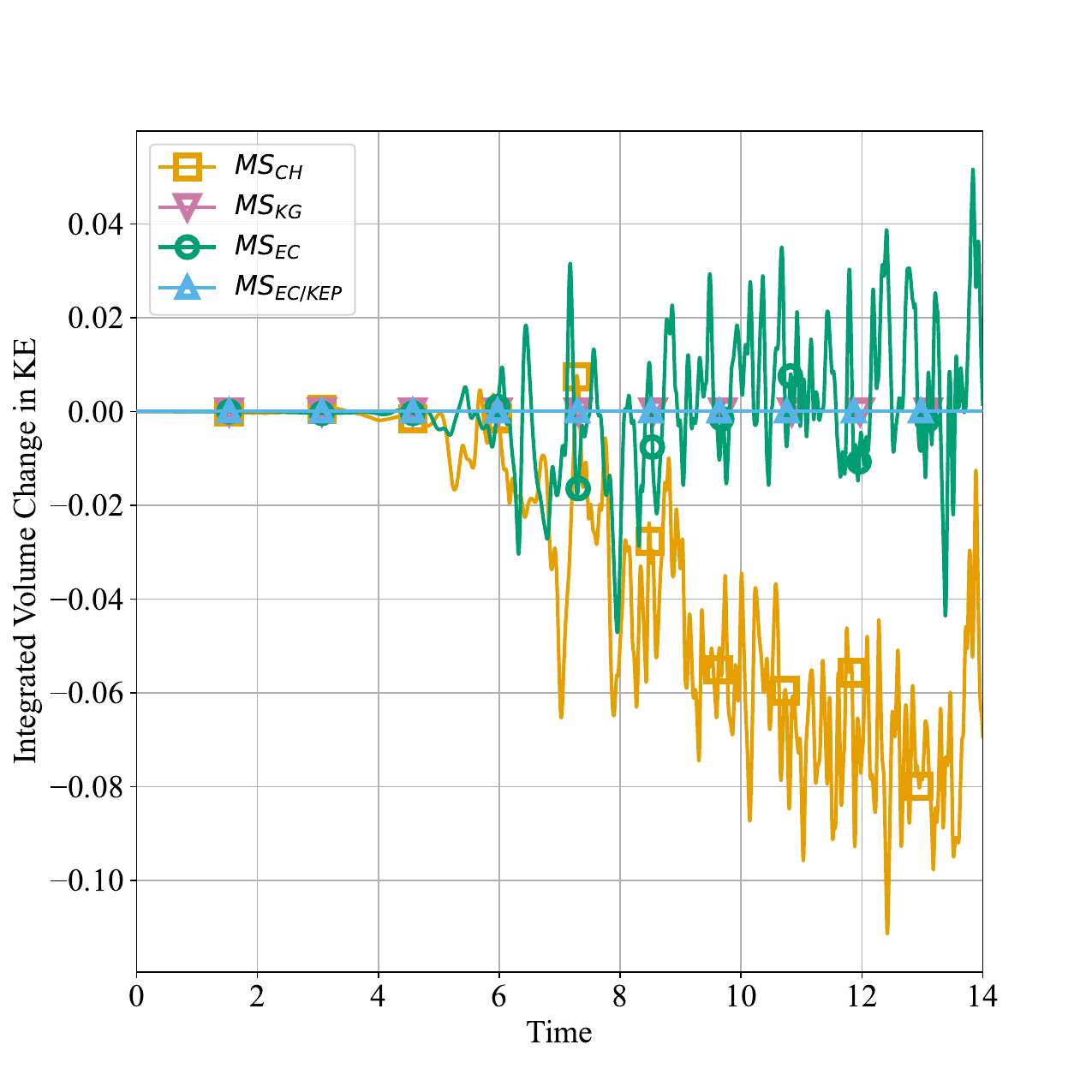}
    \includegraphics[width=0.49\linewidth]{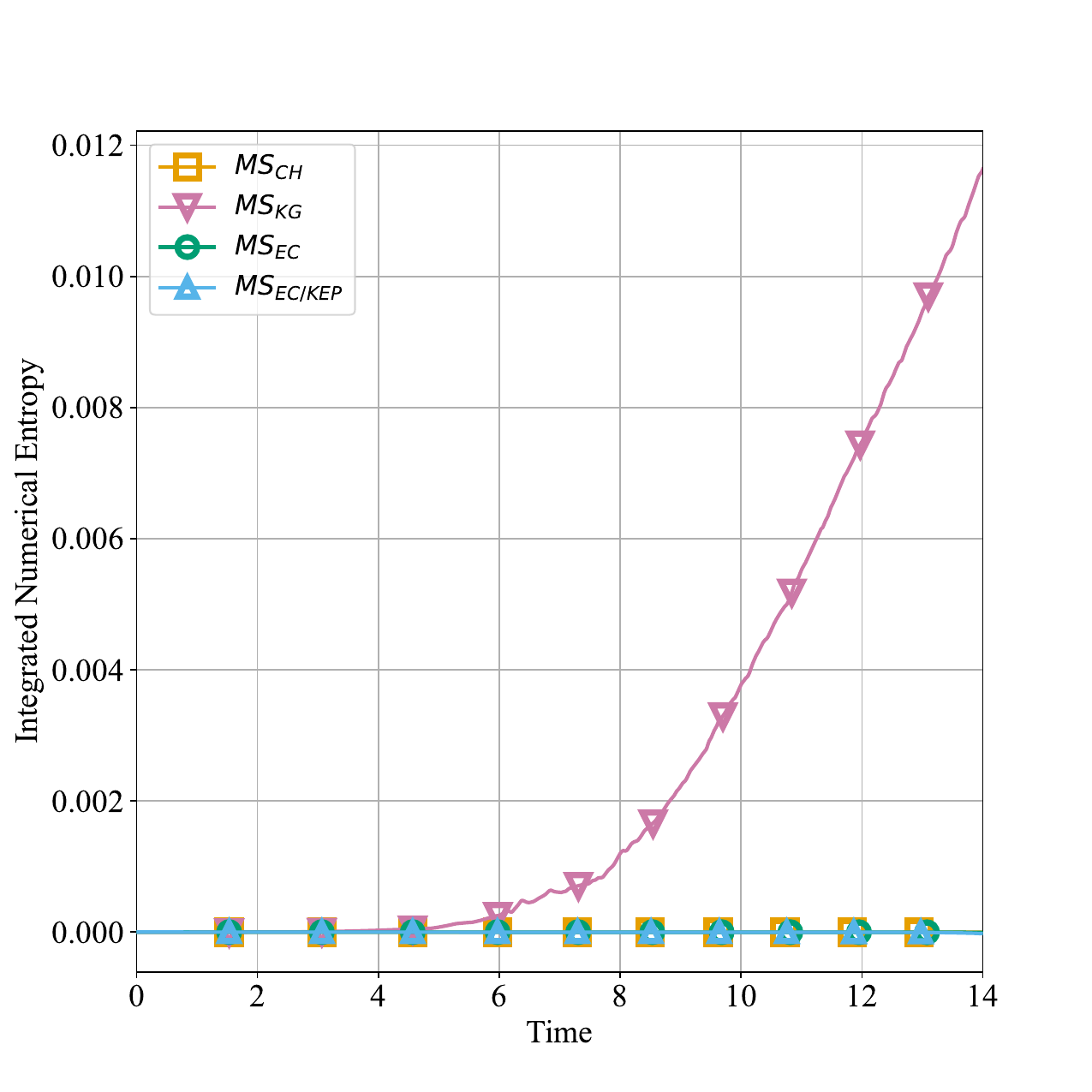}
    \caption{\textit{[Inviscid Taylor-Green Vortex]} This plot shows the results obtained for the inviscid TGV case using $p=3$, an $8^3$ grid, GLL nodes and CFL$=0.1$. We compare the $\text{MS}_{\text{CH}}$, $\text{MS}_{\text{KG}}$, $\text{MS}_{\text{EC}}$ and the newly proposed $\text{MS}_{\text{EC/KEP}}$ flux. The plot on the left shows the volume change in kinetic energy without pressure work, and the plot on the right shows the integrated entropy over time.}
    \label{fig:tgv-all-fluxes}
\end{figure}

In the interest of doing a thorough and challenging analysis of all the fluxes, we employ the inviscid Taylor-Green vortex test case similar to \cite{gassner2016split}. The test is run with $O_2$ and $N_2$ in the domain $[0,2\pi]^3$ with an $8^3$ grid and periodic boundary conditions. The flow is initialized as:
\begin{align}
    \rho &= 1.0\\
    u &= \sin{(x)}\cos{(y)}\cos{(z)}\\
    v &= -\cos{(x)}\sin{(y)}\cos{(z)}\\
    P &= \frac{1.0}{\gamma M_{inf}^2} + \frac{1.0}{16.0}(\cos{(2x}) + \cos{(2y}) + \cos{(2z}) + 2.0)\\
    Y_{O2} &= 0.5
\end{align}
The results obtained using $p3$, GLL and CFL $= 0.1$ are shown for $\text{MS}_{\text{CH}}$, $\text{MS}_{\text{KG}}$ and $\text{MS}_{\text{EC}}$ flux in Fig.~\ref{fig:tgv-ch-ir-kg-only}. We know that $\text{MS}_{\text{KG}}$ is not EC, while $\text{MS}_{\text{CH}}$ and $\text{MS}_{\text{EC}}$ are not KEP. The results for the newly proposed $\text{MS}_{\text{EC/KEP}}$ flux are shown in Fig.~\ref{fig:tgv-all-fluxes}. We see from the integrated kinetic energy and integrated entropy plots that the new flux is both EC and KEP. 

\begin{figure}[h!]
    \centering
    \includegraphics[width=0.49\linewidth]{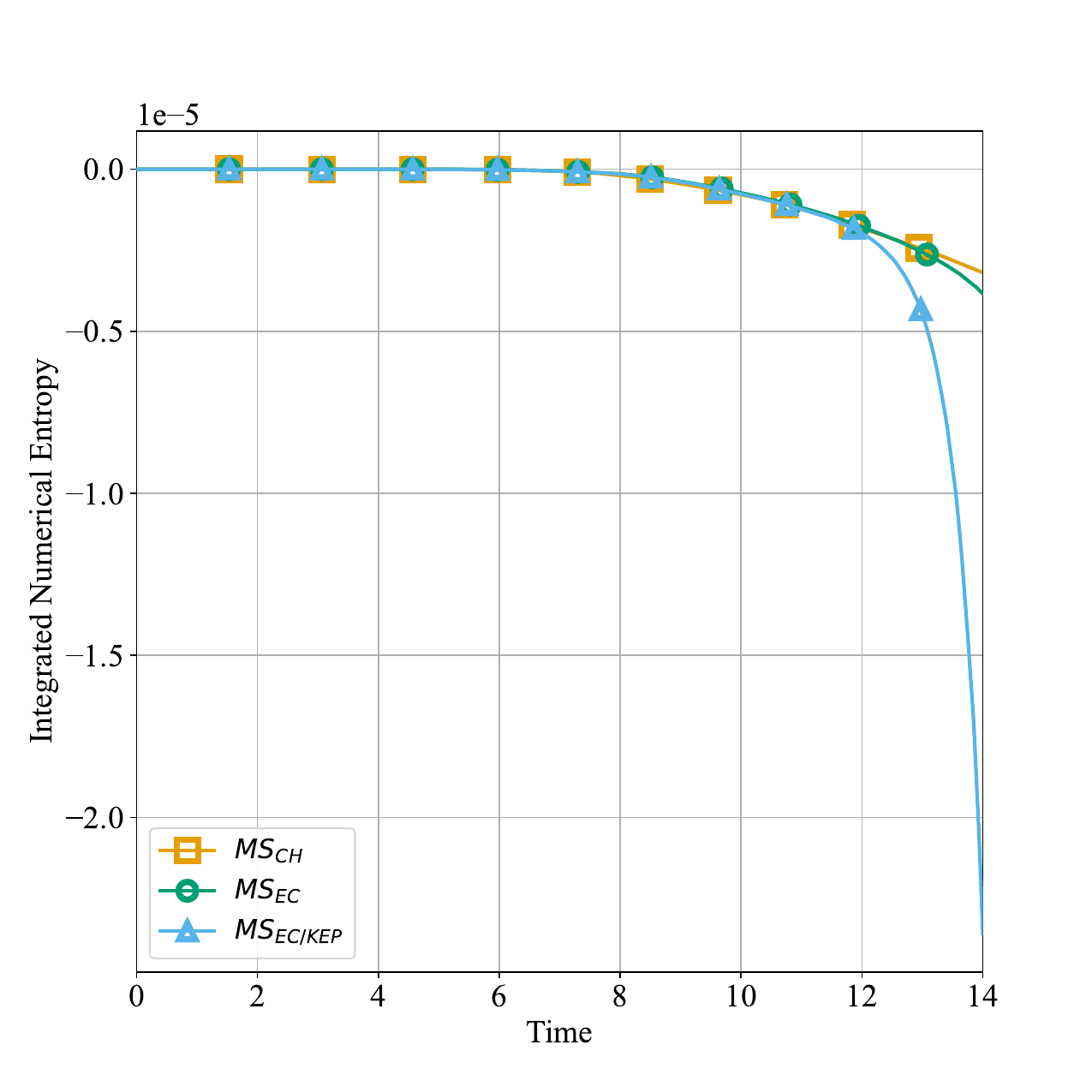}
    \includegraphics[width=0.49\linewidth]{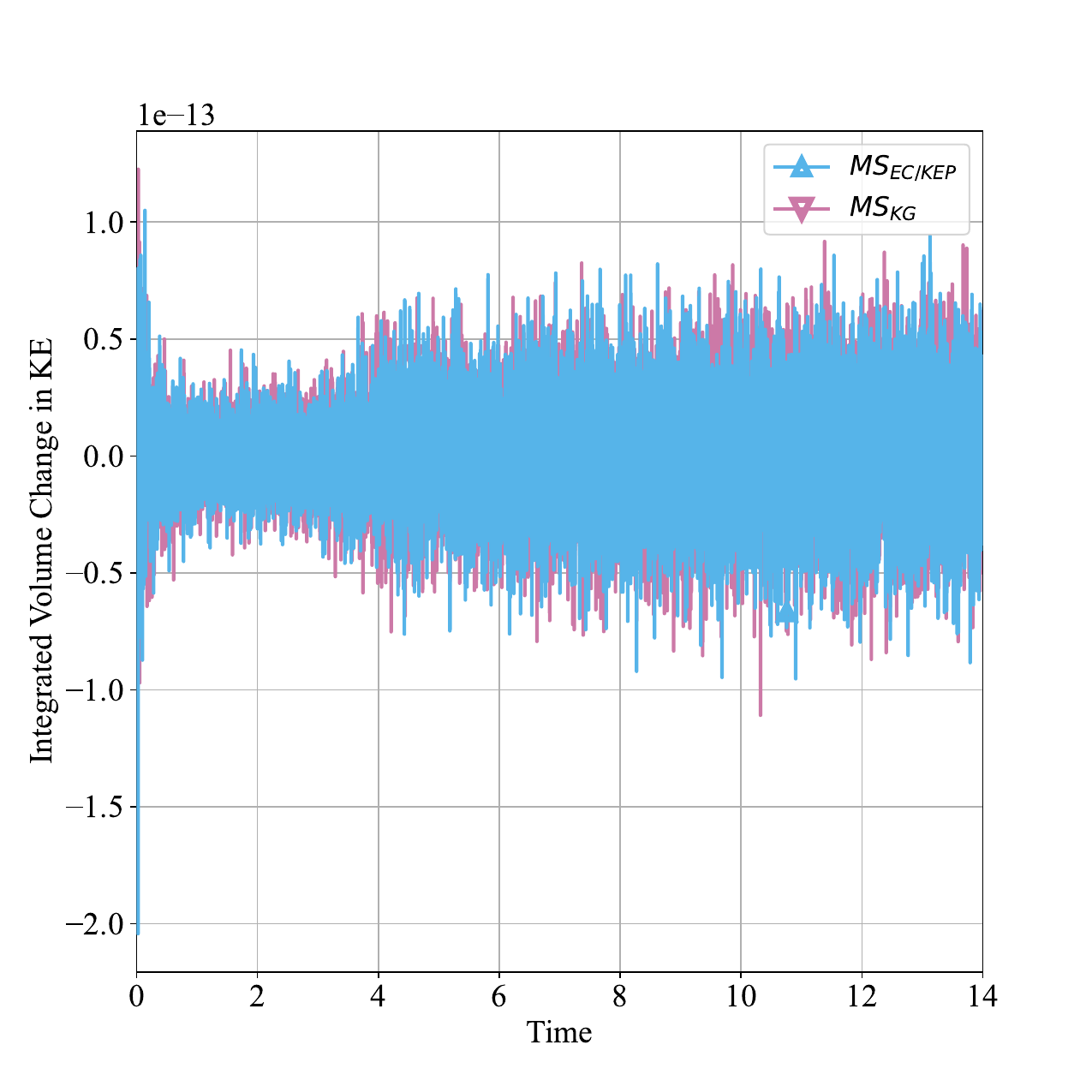}
    \caption{\textit{[Inviscid Taylor-Green Vortex]} This plot shows the results obtained for the inviscid TGV case using $p=3$, an $8^3$ grid, GLL nodes and CFL$=0.1$. We compare the change in integrated entropy over time for $CH$, $\text{MS}_{\text{EC}}$ and $\text{MS}_{\text{EC/KEP}}$ on the right and the volume change in kinetic energy without pressure work for $KG$ and $\text{MS}_{\text{EC/KEP}}$ on the left.}
    \label{fig:tgv-ra-compare}
\end{figure}

\begin{figure}
    \centering
    \includegraphics[width=0.49\linewidth]{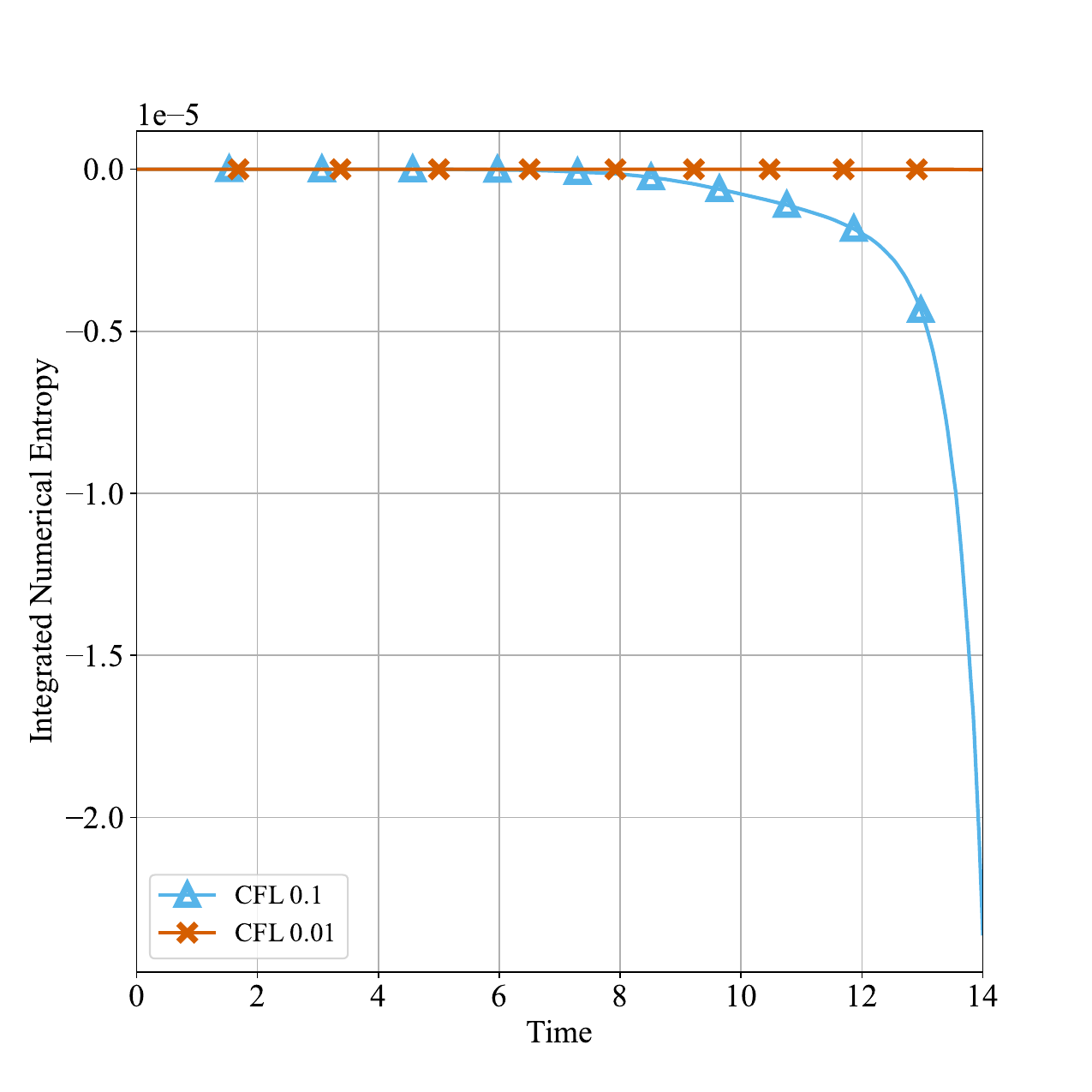}
    \includegraphics[width=0.49\linewidth]{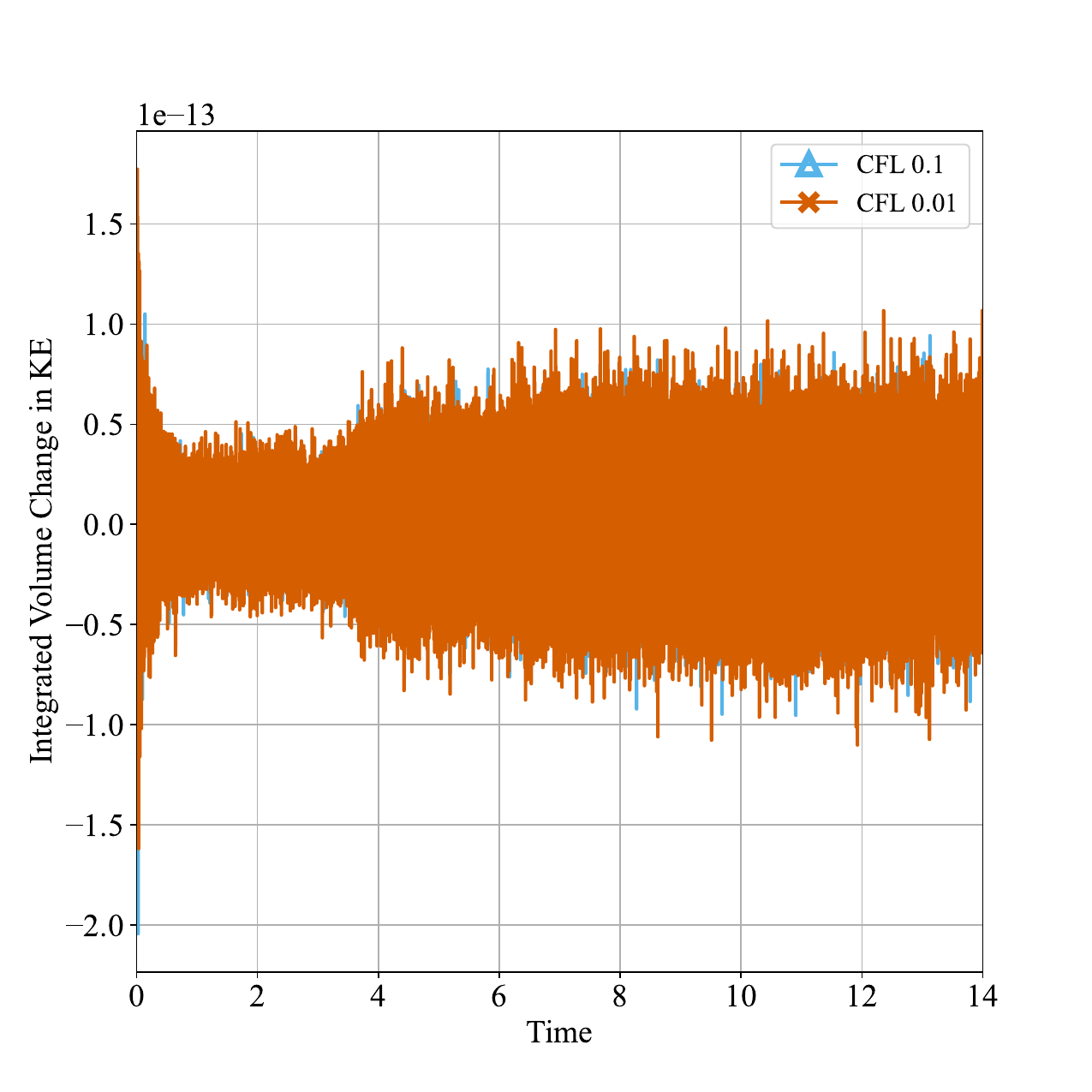}
    \caption{\textit{[Inviscid Taylor-Green Vortex]} This plot shows the results obtained for the inviscid TGV case using $p=3$, $8^3$ grid, GLL nodes and the $\text{MS}_{\text{EC/KEP}}$ flux. We plot the volume change in kinetic energy without pressure work on the left and the change in integrated entropy over time on the right for CFL$=0.1$ and CFL$=0.01$.}
    \label{fig:tgv-ra-only-time-conv}
\end{figure}

For further analysis of $\text{MS}_{\text{EC/KEP}}$, we present plots in Fig.~\ref{fig:tgv-ra-compare} where it is compared to fluxes that also possess the property that is being verified (i.e., for EC it is compared with $\text{MS}_{\text{CH}}$ and $\text{MS}_{\text{EC}}$, for KEP it is compared with $\text{MS}_{\text{KG}}$). We see that $\text{MS}_{\text{KG}}$ and $\text{MS}_{\text{EC/KEP}}$ conserve KE at machine precision. For entropy conservation we see that $\text{MS}_{\text{CH}}$ and $\text{MS}_{\text{EC/KEP}}$ are not at machine precision. To further verify that the flux is EC, we lower the CFL and compare the integrated numerical entropy. The results for $\text{MS}_{\text{EC/KEP}}$ using CFL$=0.1$ and CFL$=0.01$ are shown in Fig.~\ref{fig:tgv-ra-only-time-conv}. We observe that KE continues to be at machine precision. We also observe that the numerically integrated entropy is several orders of magnitude lower when the timestep is refined. This demonstrates that the flux is indeed EC at design order.

% \begin{figure}
%     \centering
%     \includegraphics[width=0.49\linewidth]{ms-tgv-ec-ra-lower-cfl.pdf}
%     \includegraphics[width=0.49\linewidth]{ms-tgv-kep-ra-lower-cfl.pdf}
%     \caption{\textbf{\textit{[Inviscid Taylor-Green Vortex]} $p=3$, $8^3$ grid, GLL nodes, $\text{MS}_{\text{EC/KEP}}$ flux, CFL$=0.01$. Included to show Prof the values.}}
%     \label{fig:tgv-ra-only-time-conv}
% \end{figure}
\section{Conclusion}\label{sec: conclusion}
In this short communication, we derived an entropy conservative, kinetic energy preserving two-point flux for multi-species compressible flow using the strategy of \citet{ranocha2018generalised}. {We showed that the fix for KEP is identical to the fluxes derived for the calorically perfect single-species formulation seen in \cite{ranocha2018generalised} and the thermally perfect single-species formulation seen in \cite{aiello2026formulation}. }Using the nonlinearly-stable flux reconstruction scheme, we then demonstrated the performance of the newly proposed two-point flux in comparison to the existing options. The density pulse test case shown in Sec.~\ref{sec: density-pulse} shows that the new $MS_{EC/KEP}$ flux achieves the design order of convergence and has the same magnitude of error as the Chandrashekar ($CH$) flux and the $MS_{EC}$ flux derived in this work. It also has a lower magnitude of error for pressure than the Kennedy-Gruber ($KG$) flux. It is also proven that the flux is entropy conservative through a timestep convergence study. The inviscid Taylor-Green vortex test case presented in Sec.~\ref{sec: tgv} shows that the new flux is both entropy conserving and kinetic energy preserving, while the other two-point fluxes only have one of the two properties. Since the two-point flux dictates the properties of the resulting scheme, the proposed flux is the ideal option for simulating multi-species compressible flow.
\section*{Acknowledgments}
\indent Sai Shruthi Srinivasan thanks the Vadasz Family Foundation and McGill Engineering Doctoral Award (MEDA). 

\bibliographystyle{model1-num-names}
\bibliography{refs}

@string{jsc = {J. Sci. Comput.}}

@article{tadmor1987numerical,
  title={{The numerical viscosity of entropy stable schemes for systems of conservation laws. I}},
  author={Tadmor, Eitan},
  journal={Mathematics of Computation},
  volume={49},
  number={179},
  pages={91--103},
  year={1987}
}

@article{jameson2008formulation,
  title={{Formulation of kinetic energy preserving conservative schemes for gas dynamics and direct numerical simulation of one-dimensional viscous compressible flow in a shock tube using entropy and kinetic energy preserving schemes}},
  author={Jameson, Antony},
  journal=jsc,
  volume={34},
  number={2},
  pages={188--208},
  year={2008},
  publisher={Springer}
}

@article{cicchino2025discretely,
  title={{Discretely nonlinearly stable weight-adjusted flux reconstruction high-order method for compressible flows on curvilinear grids}},
  author={Cicchino, Alexander and Nadarajah, Siva},
  journal={Journal of Computational Physics},
  volume={521},
  pages={113532},
  year={2025},
  publisher={Elsevier}
}

@article{gassner2016split,
  title={{Split form nodal discontinuous Galerkin schemes with summation-by-parts property for the compressible Euler equations}},
  author={Gassner, Gregor J and Winters, Andrew R and Kopriva, David A},
  journal={Journal of Computational Physics},
  volume={327},
  pages={39--66},
  year={2016},
  publisher={Elsevier}
}

@phdthesis{ranocha2018generalised,
  title={{Generalised summation-by-parts operators and entropy stability of numerical methods for hyperbolic balance laws}},
  author={Ranocha, Hendrik},
  year={2018},
  publisher={Cuvillier Verlag},
  school={TU Braunschweig}
}

@article{ching2024positivity,
  title={{Positivity-preserving and entropy-bounded discontinuous Galerkin method for the chemically reacting, compressible Euler equations. Part I: The one-dimensional case}},
  author={Ching, Eric J and Johnson, Ryan F and Kercher, Andrew D},
  journal={Journal of Computational Physics},
  volume={505},
  pages={112881},
  year={2024},
  publisher={Elsevier}
}

@phdthesis{gouasmi2020contributions,
  title={{Contributions to the Development of Entropy-Stable Schemes for Compressible Flows}},
  author={Gouasmi, Ayoub},
  year={2020}
}

@article{keeton2025discontinuous,
  title={{A discontinuous Galerkin spectral element method for compressible reacting flows}},
  author={Keeton, Benjamin W and Ameen, Muhsin and Pal, Pinaki},
  journal={Computer Methods in Applied Mechanics and Engineering},
  volume={446},
  pages={118277},
  year={2025},
  publisher={Elsevier}
}

@article{wang2019partial,
  title={{Partial characteristic decomposition for multi-species Euler equations}},
  author={Wang, Jian-Hang and Pan, Shucheng and Hu, Xiangyu Y and Adams, Nikolaus A},
  journal={Computers \& Fluids},
  volume={181},
  pages={364--382},
  year={2019},
  publisher={Elsevier}
}

@article{renac2021entropy,
  title={{Entropy stable, robust and high-order DGSEM for the compressible multicomponent {E}uler equations}},
  author={Renac, Florent},
  journal={Journal of Computational Physics},
  volume={445},
  pages={110584},
  year={2021},
  publisher={Elsevier}
}

@article{peyvan2023high,
  title={{High-order methods for hypersonic flows with strong shocks and real chemistry}},
  author={Peyvan, Ahmad and Shukla, Khemraj and Chan, Jesse and Karniadakis, George},
  journal={Journal of Computational Physics},
  volume={490},
  pages={112310},
  year={2023},
  publisher={Elsevier}
}

@article{badrkhani2026entropy,
  title={{An Entropy-Stable/Double-Flux scheme for the multi-component compressible Navier-Stokes equations}},
  author={Badrkhani, Vahid and Karpowski, T Jeremy P and Hasse, Christian},
  journal={Journal of Computational Physics},
  pages={115079},
  year={2026},
  publisher={Elsevier}
}

@article{ismail2009affordable,
  title={{Affordable, entropy-consistent Euler flux functions II: Entropy production at shocks}},
  author={Ismail, Farzad and Roe, Philip L},
  journal={Journal of Computational Physics},
  volume={228},
  number={15},
  pages={5410--5436},
  year={2009},
  publisher={Elsevier}
}

@article{ma2017entropy,
  title={An entropy-stable hybrid scheme for simulations of transcritical real-fluid flows},
  author={Ma, Peter C and Lv, Yu and Ihme, Matthias},
  journal={Journal of Computational Physics},
  volume={340},
  pages={330--357},
  year={2017},
  publisher={Elsevier}
}

@article{giovangigli2012multicomponent,
  title={Multicomponent flow modeling},
  author={Giovangigli, Vincent},
  journal={Science China Mathematics},
  volume={55},
  number={2},
  pages={285--308},
  year={2012},
  publisher={Springer}
}

@article{cicchino2022nonlinearly,
  title={Nonlinearly stable flux reconstruction high-order methods in split form},
  author={Cicchino, Alexander and Nadarajah, Siva and Fern{\'a}ndez, David C Del Rey},
  journal={Journal of Computational Physics},
  volume={458},
  pages={111094},
  year={2022},
  publisher={Elsevier}
}

@article{harten1983symmetric,
  title={On the symmetric form of systems of conservation laws with entropy},
  author={Harten, Amiram},
  journal={Journal of computational physics},
  volume={49},
  year={1983}
}

@article{kennedy2008reduced,
  title={Reduced aliasing formulations of the convective terms within the Navier--Stokes equations for a compressible fluid},
  author={Kennedy, Christopher A and Gruber, Andrea},
  journal={Journal of Computational Physics},
  volume={227},
  number={3},
  pages={1676--1700},
  year={2008},
  publisher={Elsevier}
}

@article{derigs2017novel,
  title={A novel averaging technique for discrete entropy-stable dissipation operators for ideal MHD},
  author={Derigs, Dominik and Winters, Andrew R and Gassner, Gregor J and Walch, Stefanie},
  journal={Journal of Computational Physics},
  volume={330},
  pages={624--632},
  year={2017},
  publisher={Elsevier}
}

@article{aiello2026formulation,
  title={Formulation of entropy-conservative discretizations for compressible flows of thermally perfect gases},
  author={Aiello, Alessandro and De Michele, Carlo and Coppola, Gennaro},
  journal={Journal of Computational Physics},
  pages={115053},
  year={2026},
  publisher={Elsevier}
}

@article{chan2026nodal,
  title={Nodal discontinuous Galerkin methods for non-ideal equations of state: pressure equilibrium preservation and entropy correction},
  author={Chan, Jesse and Ranocha, Hendrik and Park, Raymond and Lampert, Joshua and Ching, Eric and Edoh, Ayaboe},
  journal={arXiv preprint arXiv:2608.14506},
  year={2026}
}

@article{coppola2026pressure,
  title={Pressure-equilibrium-preserving and fully conservative discretization of compressible flow equations for real and thermally perfect gases},
  author={Coppola, Gennaro and Aiello, Alessandro and De Michele, Carlo},
  journal={arXiv preprint arXiv:2605.03617},
  year={2026}
}

\end{document}